\documentclass[12pt]{article}
\begin{document}
\begin{center}
{\Large\bf \boldmath  To the problem of the intrinsic magnetism in carbon-based systems: pro et contra}\footnote{Report  presented at 
the International Conference "DUBNA-NANO 2012",  
Dubna 9 - 14 July 2012, Russia.\\} \\ 
 
\vspace*{6mm}
{A. L. Kuzemsky   }\\      
{\small \it Bogoliubov Laboratory of Theoretical Physics,}\\
{\small \it  Joint Institute for Nuclear Research, Dubna, Russia.\\
kuzemsky@theor.jinr.ru;   http://theor.jinr.ru/\symbol{126}kuzemsky}         
\end{center}

\vspace*{6mm}

\begin{abstract}
The arguments supporting the existence of the intrinsic magnetism in carbon-based materials including pure 
graphene were analyzed critically together with the numerous experimental evidences denying the magnetism in
these materials. The crucial experiment of Sepioni et al~\cite{sep10} showed clearly  that no
ferromagnetism was detected in  pure graphene  at any temperature down to $2 \, K$. Neither do they found strong paramagnetism
expected due to the massive amount of edge defects. Rather, graphene is strongly diamagnetic, similar to
graphite. Thus the possible traces of a quasi-magnetic behavior which some authors observed in their samples
may be attributed rather to induced magnetism due to the impurities, defects, etc.
On the basis of the present analysis the conclusion was made that the thorough and detailed 
experimental studies of these problems only may shed light on the very complicated problem 
of the magnetism of carbon-based materials.\\

\noindent \textbf{Keywords}: 
Carbon-based materials; pure graphene; the magnetism of carbon-based materials;
intrinsic magnetism; induced magnetism; ferromagnetism, paramagnetism, diamagnetism; quasi-magnetic behavior; 
quantum theory of magnetism; the role of the low dimensionality.\\ 

\end{abstract}

\vspace*{6mm}

Many new growth points in magnetism have appeared in the last decades. 
Magnetic materials,  as a rule,   can be metals, semiconductors or insulators
which contain the ions of the transition metals or rare-earth metals with unfilled shells. 
According to Pauli exclusion principle
the electrons with parallel spins tend to avoid each other spatially.
One can say that the Pauli exclusion principle
 lies in the foundation of the quantum theory of
magnetic phenomena. 
It is worth noting that the magnetically active electrons which form the magnetic
moment can be localized or itinerant (collectivized).\\
The search for macroscopic magnetic ordering in exotic materials  and artificial devices has
attracted big attention~\cite{rod06,sieg06,stef08,coe09,spald11,stef12}.
In particular, the carbon-based materials were pushed into the first row of researches. 
The understanding and control of the magnetic properties of carbon-based materials may be of fundamental
relevance in applications in nano- and biosciences.\\
Carbon materials are unique in many ways~\cite{carb1,carb2}. They are characterized by the 
various allotropic forms that carbon materials can assume~\cite{carb2}, including the graphene - a monolayer of carbon atoms
densely packed in a honeycomb lattice~\cite{wsnano11,katz12}. It was conjectured that in addition to its transport properties a rich variety of magnetic
behavior may be expected in graphene, including even a kind of intrinsic ferromagnetism. Some hypothesis were claimed  
that connected possible spin-ordering effects with the low-dimensionality and Dirac-like electron spectrum of graphene,
thus inspiring a new kind of   magnetism  without magnetic ions.\\ 
In spite of the fact that magnetism is not usually expected in simple $s-p$ oxides like $MgO$ or in carbons like graphite 
it was supposed that  basic intrinsic defects in these systems~\cite{ston10,yazy10,he12} can be magnetic in ways 
that seem to be shared by more complex oxides. 
Moreover, a "room-temperature ferromagnetism of graphene" was claimed~\cite{wang09}. However, the mechanism of 
ferromagnetism in carbon-based materials, which contain only $s$ and $p$ electrons in
contrast to traditional ferromagnets based on $3d$ or $4f$ electrons, is rather unclear.\\ 
There are many examples of physical systems with a permanent magnetic moment in the ground state~\cite{kor07}. These systems 
are the atoms, molecules and ions with an odd number of electrons, some molecules with an even number of 
electrons ($O_{2}$  and some organic compounds) and atoms (ions) with an unfilled ($3d-$, $4f-$, $5f-$) shells.
Strong magnetic materials as a rule include $3d-$ ions with unfilled shells~\cite{kuz09}. 
(For example, $T_{C}(Fe) \approx 1040 K$  whereas $T_{C}(Gd) \approx 290 K$).\\
Thus the natural question arises: can carbone-based materials be magnetic in principle and what is the mechanism of the appearing of the
magnetic state from the point of view of the quantum theory of magnetism?\\
Previously, the "hybrid materials" known as molecular ferromagnets in which organic groups are combined with transition metal 
ions were prepared~\cite{veci01,blund04}. Here the organic groups were themselves not magnetic but were used to mediate the 
magnetism between transition metal ions. Organic ferromagnetism was first achieved using organic radicals called nitronyl nitroxides. Many organic radicals exist which have 
unpaired spins, but few are chemically stable enough to assemble into crystalline structures.
Nevertheless, it was recognized that "reports of weak magnetization in organic materials have often proved to be wrong".\\
Magnetism in carbon allotropes has indeed been a fundamental and also controversial problem for a long time~\cite{carb1,carb2}.  
Conventional wisdom has it that
carbon (containing only $s$ and $p$ electrons) does not have a spontaneous magnetic moment in any of its allotropes.
In spite of the fact that carbon is diamagnetic ($\chi \sim -6 \cdot 10^{-6}$),
in 2001 an "observation of strong magnetic signals in rhombohedral pristine $C^{60}$, indicating a Curie
temperature $T_{C}$  near 400-500K" was reported~\cite{mak1,esq1}. 
Polymerization of fullerenes can be realized through high-pressure and temperature treatments and through 
irradiation with UV light.  It can also occur through reactions with alkali metals.
Wood \emph{et al} described the "ferromagnetic fullerene"~\cite{wood02}.
This new magnetic forms of $C^{60}$ have been identified with the state which occur in the
rhombohedral polymer phase. The existence of previously reported
ferromagnetic rhombohedral $C^{60}$ was confirmed. This property has been shown to
occur over a range of preparation temperatures at $9 \, \textrm{GPa}$. The structure was shown
to be crystalline in nature containing whole undamaged buckyballs. Formation
of radicals is most likely due to thermally activated shearing of the bridging bond
resulting in dangling bond formation. With increasing temperatures this process
occurs in great enough numbers to trigger cage collapse and graphitization.
The magnetically strongest sample was formed at 800 K, and has a saturated
magnetization at 10 K, in fields above $3 \, \textrm{kOe}$, of $0.045 \,  \textrm{emu g.}$\\
Moreover, in paper~\cite{nar03} the observation 
of the ferromagnetically ordered state in a material obtained by high-pressure high-temperature treatment of the 
fullerene $C^{60}$ was confirmed. It had a saturation 
magnetization more than four times larger than that reported previously. From their data  the considerably 
higher value of $T_{C} \approx $ 820 K was estimated~\cite{nar03}.\\
The  widely advertized "discovery" of a ferromagnetic form of carbon~\cite{mak1,esq1,wood02}  stimulated   huge stream 
of  the investigations  of the carbon-based materials~\cite{mak1,esq1,wood02,nar03,han03,mak04,esq2,esq05,mak06,rao12}.
However, difficulties to reproduce those results and the unclear role of impurities casted doubts on the existence
of a ferromagnetic form of carbon.\\
And nevertheless, it was claimed that "the existence of carbon-based magnetic material requires a root-and-branch rework of magnetic theory". 
Moreover, "the existing theory for magnetism in elements with only $s$ and $p$ electron orbits (such as carbon)" should be
reconsidered in the light of the fact that there are many 
publications "describing ferromagnetic structures containing either pure carbon or carbon combined with first row elements",
in spite of "these reports were difficult to reproduce". \\
It is worth mentioning that in the publications~\cite{mak1,nar03,han03} the  characterization of the samples was not
made properly. This fact was recognized by the authors themselves~\cite{khan03,spem03,mak2}.
In paper~\cite{spem03} 
a $C^{60}$ polymer has been characterized for the first time with respect to impurity content and ferromagnetic properties 
by laterally resolved particle induced X-ray emission   and magnetic force microscopy   in order to prove the 
existence of intrinsic ferromagnetism in this material. In the sample studied the main ferromagnetic impurity found was 
iron with remarkable concentration. In spite of that fact authors insisted that they were able
"to separate between the intrinsic and extrinsic magnetic regions and to directly prove that intrinsic 
ferromagnetism exists in a $C^{60}$ polymer".\\
In 2004  the band structure calculations of rhombohedral $C^{60}$ performed in the local-spin-density approximation 
were presented~\cite{katz04}. Rhombohedral $C^{60}(Rh-C^{60})$ is a 
two-dimensional polymer of $C^{60}$ with trigonal topology. No magnetic solution exists for $Rh-C60$ and energy bands 
with different spins were found 
to be identical and not split. The calculated $C$ $2p$ partial density of states was compared to carbon K-edge X-ray 
emission and absorption spectra and showed good agreement. It was concluded that the rhombohedral distortion of $C^{60}$ 
itself cannot induce magnetic ordering in the molecular carbon. 
The result of magnetization measurements performed on the same $Rh-C^{60}$ sample corroborates this conclusion.
It is worth noting that in majority publications on the possible ferromagnetism  of carbon-based "magnetic" material
the effects of the low dimensionality~\cite{may09,kuz10} on the possible magnetic ordering were practically ignored.\\
In 2006 a retraction letter~\cite{mak3} has been published.
Some of the authors (two of them decline to sign this retraction) 
recognized that reported high-temperature ferromagnetism in a
polymeric phase of pure carbon that was purportedly free of
ferromagnetic impurities was an artefact. Other measurements
made on the same and similar samples using particle-induced
X-ray emission (PIXE) with a proton microbeam have indicated
that these had considerable iron content. Also, polymerized $C^{60}$
samples mixed with iron before polymerization had a similar Curie
temperature (500 K) to those they described~\cite{mak1}, owing to the presence of
the compound $Fe_{3}C$ (cementite). In addition, it has since been
shown that the pure rhombohedral $C^{60}$ phase is not ferromagnetic~\cite{katz04}.
Nevertheless, they concluded that "magnetic order in impurity-free graphitic structures
at room temperature has been demonstrated independently (before
and after publication of ref.~\cite{mak1}). Ferromagnetic properties may yet be
found in polymerized states of $C^{60}$ with different structural defects
and light-element $(H, O, B, N)$ content".\\
In spite of this dramatic development, the search  for magnetic order at room temperature in a system without the usual $3d$ metallic magnetic 
elements continues. It was conjectured that the 
graphite structure with defects and/or hydrogen appears to be one of the most promising candidates to find this phenomenon.\\
The irradiation effects  for the properties of carbon-based materials were found substantial.
Some evidence that proton irradiation on highly oriented pyrolytic
graphite samples may triggers ferro- or ferrimagnetism was reported~\cite{esq03}. The possibility of a
magnetism in graphene nanoislands was speculated and a defective graphene phase predicted to be
a room temperature ferromagnetic semiconductor was conjectured as well.\\
In paper~\cite{rode09}
vacancies and vacancy clusters produced by carbon ion implantation in highly oriented
pyrolytic graphite, and their annealing behavior associated with the ferromagnetism of
the implanted sample were studied using positron annihilation in conjunction with ferromagnetic
moment measurements using a superconducting quantum interferometer device
magnetometer. Author's results give some indication that the "magnetic moments" may be correlated to
the existence of the vacancy defects in the samples and this is supported by theoretical calculations
using density functional theory. The possible mechanism of magnetic order in the implanted sample was discussed.
Authors~\cite{rode09} claimed that "it has become evident \ldots that even pure carbon can show substantial paramagnetism
and even ferromagnetism".\\
In paper~\cite{esq10} recently obtained data
were discussed  using different experimental methods including magnetoresistance measurements that indicate the 
existence of metal-free high-temperature magnetic order in graphite. Intrinsic as well as extrinsic difficulties to trigger magnetic 
order by irradiation of graphite were discussed. 
The introduction of defects in the graphite structure by irradiation may be in principle a  relevant 
method to test any possible magnetic order in carbon since it allows to minimize sample handling and to estimate 
quantitatively the produced defect density in the structure. The main magnetic effects produced by proton irradiation 
have been reproduced in various further studies. X-ray magnetic circular dichroism (XMCD) studies on proton-irradiated 
spots on carbon films confirmed that the magnetic order is correlated 
to the $\pi$-electrons of carbon only, ruling out the existence of magnetic impurity contributions.
The role of defects and vacancies  continues to be  under current intensive study.\\
In paper~\cite{he12}, by means of
near-edge x-ray-absorption fine-structure (NEXAFS) and bulk magnetization measurements, it was demonstrated
that the origin of ferromagnetism in $^12C^+$ ion implanted highly oriented pyrolytic graphite (HOPG) is closely
correlated with the defect electronic states near the Fermi level. The angle-dependent NEXAFS spectra imply
that these defect-induced electronic states are extended on the graphite basal plane. It was concluded that  the origin
of electronic states to the vacancy defects created under $^{12}C^+$ ion implantation. The intensity of the observed
ferromagnetism in HOPG is sensitive to the defect density, and the narrow implantation dosage window that
produces ferromagnetism should be  optimized.\\
In paper~\cite{ugeda12},
electronic and structural characterization of divacancies in irradiated graphene was investigated.
Authors provided a thorough study of a carbon divacancy, a point defect expected to have a large impact on
the properties of graphene. Low-temperature scanning tunneling microscopy imaging of irradiated graphene
on different substrates enabled them to identify a common twofold symmetry point defect. Authors performed  
first-principles
calculations and found that the structure of this type of defect accommodates two adjacent missing atoms in a
rearranged atomic network formed by two pentagons and one octagon, with no dangling bonds. Scanning
tunneling spectroscopy measurements on divacancies generated in nearly ideal graphene showed an electronic
spectrum dominated by an empty-states resonance, which was ascribed to a nearly flat, spin-degenerated band of
$\pi$-electron nature. While the calculated electronic structure rules out the formation of a magnetic moment around
the divacancy, the generation of an electronic resonance near the Fermi level reveals divacancies as key point
defects for tuning electron transport properties in graphene systems. 
Thus the situation is still controversial~\cite{esq10}.\\
High-temperature ferromagnetism in graphene and other graphite-derived materials reported by several workers~\cite{wang09,rao12} 
has attracted considerable interest. Magnetism in graphene and graphene nanoribbons is ascribed to defects and edge states, 
the latter being an essential feature of these materials~\cite{yazy10}. Room-temperature ferromagnetism in graphene~\cite{rao12}  is 
affected by the adsorption of molecules, especially hydrogen. Inorganic graphene analogues formed by some
layered materials  also show such ferromagnetic behavior~\cite{rao12}. Magnetoresistance observed in graphene and 
graphene nanoribbons is of significance because of the potential applications.\\
The problem  of possible intrinsic magnetism of graphene-based materials  was clarified in paper~\cite{sep10}.
The authors have studied magnetization of graphene nanocrystals obtained by sonic exfoliation of graphite. No
ferromagnetism was detected at any temperature down to 2 K. Neither do they found strong paramagnetism
expected due to the massive amount of edge defects. Rather, graphene is strongly diamagnetic, similar to
graphite. Their nanocrystals exhibited only a weak paramagnetic contribution noticeable below 50 K. The
measurements yield a single species of defects responsible for the paramagnetism, with approximately
one magnetic moment per typical graphene crystallite.\\
However, the researchers~\cite{aba11} found a new way to interconnect spin and charge by applying a relatively weak magnetic 
field to graphene and found that this causes a flow of spins in the direction perpendicular to electric current, 
making a graphene sheet magnetized.
The effect resembles the one caused by spin-orbit interaction but is larger and can be tuned by varying the external 
magnetic field. They also show that graphene placed on boron nitride is an ideal material for spintronics because 
the induced magnetism extends over macroscopic distances from the current path without decay.\\

In summary, in the present work, the problem of the existence of carbon-based magnetic material
was analyzed and reconsidered to elucidate the possible relevant mechanism (if
any) which may be responsible for observed peculiarities of the "magnetic" behavior in these
systems, having in mind the quantum theory of magnetism criteria. 
The origin of magnetism lies in the orbital and spin motions of electrons and how the electrons interact 
with one another~\cite{kuz09,kuz10}. The basic object in the magnetism of condensed matter
is the magnetic moment.  The magnetic moment in practice may depend 
on the detailed environment and additional interactions such as spin-orbit, screening effects and crystal fields.
On the basis of this analysis the
conclusion was made that the thorough and detailed experimental studies of this problem only may
lead us to a better understanding of the very complicated problem of magnetism of carbon-based
materials.
%
%
%
%


\begin{thebibliography}{99}
%
%
%
\bibitem{rod06}
Roduner E 2006
\emph{Nanoscopic materials: size-dependent phenomena}   (Cambridge: Royal Society of Chemistry)
%
%
%
%
\bibitem{sieg06}
Stohr J  and  Siegmann  H C  2006  \emph{Magnetism. From Fundamentals to Nanoscale Dynamics} (Berlin: Springer) 
%
%
%
\bibitem{stef08}
Stefanita C G 2008   
\emph{From Bulk to Nano: The Many Sides of Magnetism}  
(Berlin: Springer) 
%
%
%
\bibitem{coe09}
Coey J M D  2009  \emph{Magnetism and Magnetic Materials} (Cambridge: Cambridge University Press)
%
%
%
%
\bibitem{spald11}
Spaldin N A 2011  \emph{Magnetic Materials. Fundamentals and Applications} 
2nd ed (Cambridge: Cambridge University Press)
%
%
%
%
\bibitem{stef12}
Stefanita C G 2012
\emph{Magnetism. Basics and Applications}   (Berlin: Springer)
%
%
%
%
%
%
\bibitem{carb1}
Haddon R C   1995 {\it Nature}  {\bf 378} 249
%
%
%
\bibitem{carb2}
Hirsch A   2010 {\it Nature Materials}  {\bf 9} 868
%
%
%
%
%
%
\bibitem{wsnano11}
\emph{Graphene and Its Fascinating Attributes}  2011, ed S K Pati, T Enoki  and C N R Rao 
(Singapore: World Scientific) 
%
%
%
\bibitem{katz12}
Katsnelson M I 2012
\emph{Graphene: Carbon in Two Dimensions}   (Cambridge: Cambridge University Press)
%
%
%
\bibitem{ston10}
Stoneham M 2010 {\it J. Phys. Condens. Matter}   {\bf 22}  074211
%
%
%
\bibitem{yazy10}
Yazyev O V 2010 {\it Rep. Prog. Phys.}   {\bf 73}  056501
%
%
%
\bibitem{he12}
He Zhoutong  \emph{et al.}, 2012 {\it Phys. Rev.} B {\bf 85} 144406
%
%


%
%
%
\bibitem{wang09}
Yan Wang \emph{et al}  2009 {\it Nano Letters}   {\bf 9} 220
%
%
%
%
\bibitem{kor07}
Koroleva L I and Khapaeva T M  2007 {\it Phys. Lett.} A {\bf 371} 165
%
%
%
%
%
\bibitem{kuz09}
Kuzemsky A L   2009 {\it Physics of Particles and Nuclei}  {\bf 40} 949
%
%
%
%
\bibitem{veci01}
\emph{$\pi$-Electron Magnetism: From Molecules to Magnetic Materials} 2001
(Structure and Bonding vol.100)  ed  J Veciana and  D Arcon (Berlin: Springer)
%
%
%
\bibitem{blund04}
Blundell S J and Pratt  F L   2004 {\it J. Phys.: Condens. Matter}  {\bf 16} R771
%
%
%
%
%
\bibitem{mak1}
Makarova T L \emph{et al}  2001 {\it Nature}  {\bf 413} 690
%
%
%
\bibitem{esq1}
Hohne R and Esquinazi P   2002 {\it Adv. Mater.}  {\bf 14} 753
%
%
%
\bibitem{wood02}
Wood R A   \emph{et al}  2002 {\it J. Phys.: Condens. Matter}  {\bf 14} L385
%
%
%
%
\bibitem{nar03}
Narozhnyi V N  \emph{et al} 2003 {\it Physica}  B  {\bf 329-333} 1217
%
%
%
\bibitem{han03}
Han K H \emph{et al} 2003 {\it Carbon}     {\bf 41} 785
%
%
%
%
%
\bibitem{mak04}
Makarova T L    2004 {\it Semiconductors}  {\bf 38} 615
%
%
%
%
\bibitem{esq2}
Esquinazi P \emph{et al}  2005 {\it Phase Transitions}  {\bf 78} 175
%
%
%
%
\bibitem{esq05}
Esquinazi P  and Hohne R 2005 {\it JMMM}  {\bf 290-291} 20
%
%
%
\bibitem{mak06}
\emph{Carbon-based Magnetism: An Overview of the Magnetism of Metal Free Carbon-based Compounds and Materials}  2006, 
ed T Makarova  and   F P  Parada  (Amsterdam: Elsevier) 
%
%
%
%
%
%
%
%
\bibitem{khan03}
Han K H \emph{et al} 2003 {\it Carbon}     {\bf 41} 2425
%
%
%
\bibitem{spem03}
D. Spemann \emph{et al} 2003 {\it Nuclear Instruments and Methods in Physics Research}  B  {\bf 210} 531
%
%
%
%
\bibitem{mak2}
Makarova T L \emph{et al}  2005 {\it Nature}  {\bf 436} 1200
%
%
%
\bibitem{katz04}
Boukhvalov D W \emph{et al}  2004 {\it Phys. Rev.} B   {\bf 69} 115425
%
%
%
%
%
\bibitem{may09}
Mayama H  and Naito T 2009   {\it Physica}  E  {\bf 41} 1878
%
%
%
\bibitem{kuz10}
Kuzemsky A L  2010 {\it Int. J. Mod. Phys.} B {\bf 24} 835
%
%
%
%
\bibitem{mak3}
Makarova T L \emph{et al}  2006 {\it Nature}  {\bf 440} 707
%
%
%
%
\bibitem{rode09}
Rode A V \emph{et al} 2009 {\it Carbon}     {\bf 47} 1399
%
%
%
\bibitem{esq03}
Esquinazi P \emph{et al}  2005 {\it Phys. Rev. Lett.}  {\bf 91} 227201
%
%
%
\bibitem{esq10}
Esquinazi P \emph{et al}  2010 {\it JMMM}  {\bf 322} 1156
%
%

%
%
\bibitem{ugeda12}
Ugeda M M  \emph{et al.}, 2012 {\it Phys. Rev.} B {\bf 85} 121402(R)
%
%
%
%
%
\bibitem{rao12}
Rao C N R   \emph{et al} 2012 {\it Chem. Sci.}  {\bf 3} 45
%
%
%
%
%
%
\bibitem{sep10}
Sepioni M \emph{et al}  2010 {\it Phys. Rev. Lett.}  {\bf 105} 207205
%
%
%
\bibitem{aba11}
Abanin D A   \emph{et al}  2011 {\it Science}  {\bf 332} 328  
%
%
%
%
\end{thebibliography}
\end{document}